\documentclass[prb,showpacs,preprintnumbers,amsmath,amssymb,letterpaper,twocolumn]{revtex4}

\usepackage{graphicx}
\usepackage{dcolumn}
\usepackage{bm}
\usepackage{latexsym}
\usepackage{epsfig}
\usepackage{verbatim}
\usepackage{hyperref}
\usepackage{color}

\newcommand{\be}{\begin{equation}}
\newcommand{\ee}{\end{equation}}
\newcommand{\bea}{\begin{eqnarray}}
\newcommand{\eea}{\end{eqnarray}}

\renewcommand{\Re}{\mathrm{Re}}

\newcommand{\eref}[1]{Eq.~(\ref{#1})}
\newcommand{\esref}[1]{Eqs.~(\ref{#1})}
\newcommand{\rref}[1]{(\ref{#1})}
\newcommand{\ep}{\varepsilon}
\newcommand{\ocite}[1]{Ref.~\onlinecite{#1}}

\newcommand{\cG}{\check G}
\newcommand{\eps}{\varepsilon}

\newcommand{\tz}{\hat\tau_z}
\newcommand{\ty}{\hat\tau_y}
\newcommand{\Aw}{{\bf A}_{\omega}}
\newcommand{\jw}{{\bf j}_{\omega}}
\newcommand{\hG}{\hat G}
\newcommand{\la}{\langle}
\newcommand{\ra}{\rangle}

\newcommand{\ccG}{\check{\cal G}}
\newcommand{\bp}{{\bf p}}

\newcommand{\bk}{{\bf k}}
\newcommand{\bn}{{\bf n}}

\newcommand{\exclude}[1]{}

\begin{document}

\title{Effect of Quasiparticles Injection on the AC Response of a
  Superconductor}

\author{G. Catelani}
\affiliation{Departments of Physics and Applied Physics, Yale
University, New Haven, Connecticut 06520, USA}

\author{K. E. Nagaev,}
\affiliation{Institute of Radioengineering and Electronics, Mokhovaya
  11-7, Moscow, 125009 Russia}

\author{L. I. Glazman}
\affiliation{Departments of Physics and Applied Physics, Yale
University, New Haven, Connecticut 06520, USA}

\date{\today}

\begin{abstract}
  We calculate the AC linear response of a superconductor in a
  nonequilibrium electronic state.  The nonequilibrium state is
  produced by injecting quasiparticles into the superconductor from normal leads
  through asymmetric tunnel contacts. The dissipative part of the
  response is drastically increased by the injected quasiparticles and
  is proportional to their total number regardless of the imbalance
  between the numbers of electron-like and hole-like excitations.
\end{abstract}

\pacs{74.25.N-,74.40.Gh}

\maketitle

\section{Introduction}
\label{sec:intro}
Over the years, measurement of the impedance was used extensively in
the studies of spectra of elementary charge carriers in normal metals
and superconductors. In the case of normal-metal single crystals,
measurements of surface impedance in conditions of cyclotron resonance
were instrumental in the reconstruction of the Fermi surface
geometry.\cite{Ziman} In superconductors, the temperature dependence
of the impedance allowed one to investigate the appearance of the BCS
gap in the spectrum of quasiparticles.\cite{Tinkham} Recent interest
to the AC response of superconductors is driven by its application for
studying new materials\cite{HTc,Dolgov} and by the use of well-studied
superconductors as material for high-quality resonators.  Depending on
the problem, it is important to realize the lowest-possible
dissipation rate of a resonator, or its well-defined response to a
perturbation. The former goal is central for superconducting qubit
physics\cite{qubit,qubit2} and quantum optics,\cite{quantopt,quantopt2} where one is
interested in achieving the longest-possible coherence times.  The
latter one is important, {\sl e.g.}, for the use of superconductors in
microwave kinetic inductance detectors.\cite{Gao}

To interpret the measured impedance, one has to compare it with
theoretical predictions.  However up to now, no microscopic
calculations of impedance of superconductors under nonequilibrium
conditions were available.  The frequency and temperature dependence
of the complex conductivity of a superconductor within the BCS theory
was first evaluated in the seminal paper of Mattis and
Bardeen,\cite{MatBar} where the limit of short electron mean free path
(the ``dirty superconductor'' limit) was considered. The non-local
conductivity of a clean superconductor was derived by Abrikosov,
Gorkov and Khalatnikov.\cite{AGK} Later works of Nam\cite{Nam} were
aimed at bridging the two limiting cases.  The theory developed in
Refs.~\onlinecite{MatBar,AGK,Nam} addresses superconductors at thermal
equilibrium. If desired, this condition is fairly easy to achieve at
not-too-low temperatures. Indeed, even the early measurements have
demonstrated the hallmarks of the BCS behavior of the dissipative part
of the impedance, including its thermal-activation behavior at sub-gap
frequencies.\cite{earlmeas} At the same time, experiments involving
impedance measurements in non-equilibrium conditions relied on
heuristic extension of Mattis-Bardeen formula, amounting to the
replacement of the equilibrium quasiparticle distribution function in
the formula by a non-equilibrium one.\cite{Gao,Klap}

\begin{figure}[!bt]
\begin{center}
\includegraphics[width=0.44\textwidth]{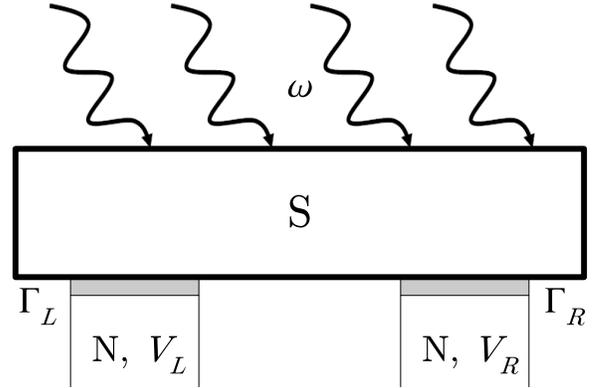}
\end{center}
\caption{N-I-S-I-N structure: the superconductor (S) is connected to two normal leads (N) -- maintained
  at different voltages (bias $V=V_R-V_L$) -- via tunnel junctions
  characterized by tunneling rates $\Gamma_R$, $\Gamma_L$. The
  superconductor is also subject to a weak electromagnetic field
  oscillating with frequency $\omega$.}
\label{fig1}
\end{figure}

The goal of this paper is
to microscopically evaluate the linear AC conductivity of a
superconductor in a concrete setup allowing a controlled
perturbation of the quasiparticle
distribution function. To this end, we analyze the steady state and
linear AC conductivity of a superconductor brought out of equilibrium
by electron tunneling through two junctions connecting the
superconductor to normal leads (N-I-S-I-N structure -- see
Fig.~\ref{fig1}); the structure is biased by a constant voltage $V$.

We find the steady-state electron distribution at finite temperature
$T$ and voltage $V$, assuming these two scales small compared to the
quasiparticle energy gap $\Delta$. In the case of unequal
conductances of the two tunnel junctions, charge imbalance is
created along with a perturbation of the energy distribution of
quasiparticles. We evaluate the complex AC conductivity of the
superconductor in this non-equilibrium state and show that only the
energy mode of the quasiparticle distribution enters in the proper
generalization of the Mattis-Bardeen formula. At $eV, k_B
T\ll\Delta$ we cast the result for conductivity in terms of $T$ and
quasiparticle density $n_{qp}$. That form extrapolates between the
equilibrium result (where $n_{qp}$ is a function of $T$ only) and
the non-equilibrium one, where $n_{qp}$ is a function of $V$ and
$T$, with arbitrary ratio $eV/k_B T$.

In the next Section, we formulate the problem in terms of matrix Green
functions. The steady state of the electrons in the superconductor
formed in the presence of finite bias applied to the N-I-S-I-N
structure is found in Section~\ref{sec:steady}, where we also
establish the correspondence between the descriptions in terms of the
matrix distribution function $\hat{F}(\varepsilon)$ and the scalar
distribution function of quasiparticles $f_\xi$ (the latter was used
in the original Mattis-Bardeen theory for the AC conductivity at
equilibrium). Using the description in terms of $f_\xi$, we include
the electron-phonon interaction into the consideration of
quasiparticle kinetics.
The AC conductivity at low temperature and bias (but at arbitrary
$eV/k_B T$) is analyzed in Section~\ref{sec:emr}.
Throughout the paper, we use units
$\hbar= k_B =1$.

\section{Electron dynamics in a superconductor subject to DC bias and
  weak AC field}
\label{sec:mod}

We consider a diffusive superconductor connected to two normal leads,
left ($L$) and right ($R$), via tunnel barriers and exposed to an
external, time-dependent electric field.  The system properties can be
described in terms of disorder-averaged matrix Green's functions for
the superconductor, $\ccG_S(\bk, t, t')$, and for the electrodes,
$\ccG_i(\bp_i, t, t')$, $i=L,R$.  Each of the matrices $\ccG$ has the
form, in Keldysh space\cite{rev}
\be
 \ccG
 =
 \left(
  \begin{array}{cc}
   \hat{\cal G}^R & \hat{\cal G}^K\\
   0     & \hat{\cal G}^A
  \end{array}
 \right).
\label{matrix}
\ee
The elements of this matrix are $2\times 2$ matrices in Nambu particle-hole space.
The superconductor Green's function obeys the Dyson equation
\begin{eqnarray}
 \left[
   i\tz\frac{\partial}{\partial t} - \left(\frac{\left(\bk-e{\bf A}(t)\right)^2}{2m} - E_F\right) + i\ty\Delta
 \right]
 \ccG_S(\bk, t, t')
\nonumber \\
- \int dt'' \frac{1}{2\pi\tau\nu_S} \sum_{\bk'} \ccG_S(\bk', t, t'') \ccG_S(\bk, t'', t')
\nonumber \\
 =\hat{1}\,\delta(t-t') + \int dt''
 \sum_{i,\bp_i} T_i^2\,\ccG_i(\bp_i, t, t'')
 \ccG_S(\bk, t'', t').\
\label{Dyson}
\end{eqnarray}
Here ${\bf A(t)}$ is the vector potential, which is related to the electric field via
${\bf E}= -\partial{\bf A}/\partial t$. The coefficient $1/\tau$ is the impurity scattering rate,
and $\nu_S$ is the density of states at the Fermi level in the superconductor.
The matrix element $T_i$ for tunneling into lead $i$ determines the dimensionless conductance
$\mathrm{g}_i=8\pi^2 \nu_S\nu_i T_i^2$ of
junction $i$, where $\nu_i$ is the density of states in the lead. The assumption
\be\label{gg-cond}
\mathrm{g}_L+\mathrm{g}_R \ll \mathrm{g}_S,
\ee
where $\mathrm{g}_S$ is the normal-state
conductance of the superconductor, justifies the use of the tunneling Hamiltonian from which the last
term in \eref{Dyson} is derived. The same assumption enables us to neglect
small spatial variations of the order parameter, which we take to be uniform, real, and time-independent.
The last two conditions amount to a choice of gauge.

Since we are interested in the linear response to the external field, we
first consider the system in the absence of field, but with the leads
at different potentials. This is the focus of the next section.

\section{Non-equilibrium steady state}
\label{sec:steady}

When there is no external field (${\bf A}=0$), the Green's functions
are isotropic in
momentum space and depend on the the difference $t-t'$. Then the
semiclassical Green's functions
\be \cG(t-t',\bn) = \frac{i}{\pi}
\int\!d\xi_{\bp}\,\ccG(\bp, t-t')\, , \quad \bn = \frac{\bp}{|\bp|},
 \label{int_xi}
 \ee
where $\xi_{\bp} = p^2/2m
 - E_F$, depend only on the time difference, and not on the momentum
 direction $\bn$. Therefore, we omit the argument $\bn$ for the rest of this section.
In the normal leads, the elements of $\cG_i$ are (in the frequency domain)
\begin{eqnarray}
\hat G_i^R & = & \tz \, , \qquad \hat G_i^A = -\tz \label{GiRA}\, ,\\
\hat G_i^K & = & \hat G_i^R \hat n_i - \hat n_i \hat G_i^A \, , \label{K_el-1}
\end{eqnarray}
with
\be
 \hat{n}_{i}(\eps)
 =
 \left(
  \begin{array}{cc}
   n(\eps-eV_{i})      & 0\\
   0        & n(\eps+eV_{i})
  \end{array}
 \right),
 \
 n(\eps)=\tanh\!\left(\frac{\eps}{2T}\right).
\label{n-def}
\ee
The potentials $V_i$ in the leads are measured from the
superconductor chemical potential.  Using \eref{GiRA}, \eref{K_el-1}
can be rewritten as
\be
  \hG^K_i = 2n^0_{i}\,\tz + 2n^1_{i}\,\hat{1},
  \label{K-el-2}
\ee
where
\be\begin{split}
 n^0_{i} & = \frac{1}{2}\,[n(\eps-eV_i) + n(\eps+eV_i)],
 \\
 n^1_{i} & = \frac{1}{2}\,[n(\eps-eV_i) - n(\eps+eV_i)].
 \label{f-def}
\end{split}\ee

The superconductor Green's function $\cG_s$ is determined by \eref{Dyson};
taking the difference between \eref{Dyson} and its conjugate, and
using \esref{int_xi}, we find
\be
 \left[ \eps\tz + i\Delta\ty , \cG_S \right]
 =-i\left[\Gamma_L\,\cG_L + \Gamma_R\,\cG_R,\cG_S\right],
 \label{basic}
\ee
where $\Gamma_{i}=\pi \nu_{i}\,T_{i}^2$ are the tunneling rates and
$\nu_i$ are the densities of states
in the leads $i=L,R$.
This equation must be supplemented by the normalization condition,\cite{rammsm}
which in the frequency domain takes the form
\be\label{normcond}
\cG_S(\ep)\cG_S(\ep) = \hat 1 \,.
\ee

Isolating the $R(A)$ component of \eref{basic} gives
\be
 \hat{H}^{R(A)}\,\hG_S^{R(A)} - \hG_S^{R(A)}\,\hat{H}^{R(A)} =0,
 \label{basic-R}
\ee
where
\be
 \hat{H}^{R(A)} = (\eps \pm i\Gamma_L \pm i\Gamma_R)\,\tz + i\Delta\ty.
 \label{HR-2}
\ee
In view of the $R(A)$ component of the normalization condition \rref{normcond},
$(\hG_S^{R(A)})^2=\hat{1}$, the solution to \eref{basic-R} can be written as
\be
 \hG_S^{R(A)}=\hat{H}^{R(A)}/\tilde\xi^{R(A)},
 \label{GR}
\ee
where
\be
 \tilde\xi^{R(A)}
 =\pm\left[
   (\eps \pm i\Gamma_L \pm i\Gamma_R)^2 - \Delta^2
  \right]^{1/2}.
 \label{xi_R}
\ee

In the limit $\Gamma_i \to 0$, \eref{GR} reduces to the well-known
semiclassical BCS expression.\cite{rammsm} When $\Gamma_i \neq 0$ the
excitations in the superconductor have finite lifetime due to
tunneling into the normal leads and this causes broadening of the
density of states.\cite{SnymanNazarov} In what follows we assume that
the broadening is much smaller than the gap, $\Gamma_L+\Gamma_R \ll
\Delta$. Since the dimensionless conductance, as defined in the text
before \eref{gg-cond}, equals $g_i = 8\pi \Gamma_i/\delta$, with
$\delta$ being the level spacing in the superconductor, we can express
this condition as
\be\label{gDd}
\mathrm{g}_L+\mathrm{g}_R\ll \Delta/\delta.
\ee

We now consider the Keldysh component of \eref{basic}. It may be written in the form
\be\begin{split}
 \hat{H}^R\,\hG^K_S - \hG^K_S\,\hat{H}^A
 = -i\Gamma_L\!
 \left(
  \hG^K_L\,\hG^A_S - \hG^R_S\,\hG^K_L
 \right)
\\ -i\Gamma_R\!
 \left(
  \hG^K_R\,\hG^A_S - \hG^R_S\,\hG^K_R
 \right).
 \label{basic-K}
\end{split}\ee
Its left-hand side can be simplified as follows: first, we use \eref{GR} to rewrite
$\hat H^{R(A)}$ in terms of $\hG^{R(A)}_S$;
then we replace the advanced
Green's function with the retarded one employing the orthogonality condition [the
Keldysh component of \eref{normcond}]
$$
 \hG^R_S\,\hG^K_S + \hG^K_S\,\hG^A_S = 0 \, .
$$
Multiplying the resulting equation by $\hG^R_S$ from the left and using again $(\hG_S^R)^2=\hat{1}$, we
arrive at
\be\begin{split}
 \left(
  \tilde{\xi}^R + \tilde{\xi}^A
 \right)
 \hG^K_S
 =i\Gamma_L\!
 \left(
  \hG^K_L - \hG^R_S\,\hG^K_L\,\hG^A_S
 \right)
 \\ +i\Gamma_R\!
 \left(
  \hG^K_R - \hG^R_S\,\hG^K_R\,\hG^A_S
 \right).
 \label{K-eq-2}
\end{split}\ee

If $|\eps|<\Delta$, the sum $\tilde{\xi}^R + \tilde{\xi}^A$ is of the order of $\Delta$, i.~e. much larger
than $\Gamma_L$ and $\Gamma_R$. However, if $|\eps| >\Delta$ this
quantity is proportional to these small parameters and can be
approximated as
\be\begin{split}
 \tilde{\xi}^R + \tilde{\xi}^A
 \approx
i(\Gamma_R + \Gamma_L)\,\eps
 \left( \frac{1}{\xi^R} - \frac{1}{\xi^A} \right),
 \label{xi_sum}
\end{split}\ee
where
\be\label{xi-def}
\xi^{R(A)} = \pm\left[(\eps\pm i0^+)-\Delta^2\right]^{1/2}.
\ee
Consistently with this approximation,
the Green's functions $G^{R/A}_S$ in the right-hand side of \eref{basic-K} should be taken at zero
order in $\Gamma_i$'s.
Using \esref{K-el-2} and \rref{GR}, after some algebra we obtain
\be\begin{split}
 \hG^K_S=\left(\frac{1}{\xi^R} - \frac{1}{\xi^A}\right)
 \Biggl[
  n^0(\eps\tz + i\Delta\ty)
  +
  n^1\frac{\eps^2 - \Delta^2}{\eps}
 \Biggr],
 \label{K-sol-3}
\end{split}\ee
where ($a=0,1$)
\be\label{fa-def}
n^a = \frac{\Gamma_L\,n^a_{L} + \Gamma_R\,n^a_{R}}{\Gamma_L + \Gamma_R}
\ee
with $n^a_{L,R}$ of \eref{f-def}.  We stress that \esref{xi_sum}-\rref{K-sol-3} are not restricted
to junctions with conductances smaller than the conductance quantum, the conditions of applicability
being given by \esref{gg-cond} and \rref{gDd}.

\begin{comment}
The Keldysh Green's function $\hG^K_S$ of \eref{K-sol-3} can be presented in terms of a
matrix distribution function $\hat F$ as
%
\be
 \hG^K_S = \hG^R_S\,\hat{F} - \hat{F}\,\hG^A_S,
 \label{K-F}
\ee
%
where
%
\be
 \hat{F}= n^0
 +\left(\tz + i\,\frac{\Delta}{\eps}\,\ty\right)
n^1 .
 \label{F-def}
\ee
%
We note that the matrix distribution function $\hat F$ is not
diagonal. This is in contrast with the usual
assumption\cite{Kopnin,LarOvc} that in non-equilibrium situations this
matrix is a linear combination of the identity matrix and $\tz$ only.
Nevertheless, there are only two independent components, since the
off-diagonal $\ty$ term is proportional to the $\tz$ one. In
Sec.~\ref{sec:qpdf} we show that this is indeed the structure to be expected for
the system under consideration and relate the two components of $\hat
F$ to the so-called energy and charge modes of the Bogoliubov
quasiparticle distribution function.
\end{comment}

The matrix $\hG^K_S$
depends on the voltages $V_L$, $V_R$ via
\esref{f-def} and \rref{fa-def}. However, under steady-state conditions these two quantities are
not independent and, as we discuss in the next section, the state of the superconductor
is fully determined by their difference $V=V_R-V_L$.

\subsection{Current and potentials}

The results of the previous sections rest on the steady-state assumption.
It means that the total current flowing out of the
superconductor must vanish. This requirement, as we show below,
defines the division of the applied bias between the two tunnel
junctions connecting the superconductor to the normal leads.

The current leaving through the left tunnel contact is given by
\be\begin{split}
 I_L =
 \frac{1}{4}\,e\nu_S\Gamma_L \int\!d\eps\,{\rm Tr}
 \bigg[
  \tz\Big(
   {\hG}^R_L\,{\hG}^K_S + {\hG}^K_L\,{\hG}^A_S
 \\  -{\hG}^R_S\,{\hG}^K_L - {\hG}^K_S\,{\hG}^A_L
  \Big)
 \bigg].
 \label{I_L-2}
\end{split}\ee
The current $I_R$ into the right contact is found by replacing $L\to R$.
Therefore the total outgoing current is
\be
 I_L + I_R = e\nu_S \int d\eps\!
 \left( \frac{1}{\xi^R} - \frac{1}{\xi^A}\right) \frac{\Delta^2}{\eps}\,
 (\Gamma_L\,n_{1L} + \Gamma_R\,n_{1R})
 \label{I_L+I_R}
\ee
and must be zero under steady-state conditions, as discussed above.
[We used Eqs.~(\ref{K-el-2}), (\ref{GR}), and (\ref{K-sol-3}) in
derivation of Eq.~(\ref{I_L+I_R}).] In other words, the requirement
\be\label{Isum-cond}
I_L+I_R=0
\ee
determines the relationship between the voltages $V_L$, $V_R$, that
enter as parameters in \eref{fa-def}.

Let us consider explicitly the regime of low temperatures and voltages
so that ${\rm max}\{eV_L,eV_R,T\}\ll\Delta$. Then the following
approximate expression for the functions $n^1_{L(R)}$ are valid at
energies $|\eps|>\Delta$:
\be
 n^1_{i}(\eps) \simeq -2\,\sinh(eV_i/T)\,\exp(-|\eps|/T) \, .
 \label{f1-exp}
\ee
Substituting into \esref{I_L+I_R}-\rref{Isum-cond} we obtain the condition
\be
 \Gamma_L\,\sinh(eV_L/T) + \Gamma_R\,\sinh(eV_R/T) =0.
 \label{cur_cons}
\ee
Introducing the bias $V=V_R-V_L$, we find from this equation that
\be
 \sinh(eV_L/T)
 =-\frac
 {\Gamma_R\,\sinh(eV/T)}
 {\sqrt{ \Gamma_L^2 + \Gamma_R^2 + 2\,\Gamma_L\,\Gamma_R\,\cosh(eV/T)}}.
 \label{sinh}
\ee
Thanks to this relation, we can reexpress voltages $V_L$, $V_R$ in terms of the bias $V$, and
the non-equilibrium steady state in the limit
$eV, T \ll \Delta$ is fully determined by these quantities.

\subsection{Quasiparticle distribution function}
\label{sec:qpdf}

In the preceding sections we have found an explicit expression for the
non-equilibrium Keldysh Green's function for the biased
superconductor, \eref{K-sol-3}.
In this section we derive the relationship between the
Keldysh Green's function
and the distribution function $f$ for
the Bogoliubov quasiparticles.

Our starting point is the definition of $\hG^K_S$ (in the time domain) in terms of creation and annihilation
operators
\begin{eqnarray}
&&\hG^K_S(t,t')
\\ &&= \frac{\tz}{\pi \nu_S} \sum_{\bp} \left(
\begin{matrix}
\langle[c_{m\uparrow}(t), c^\dagger_{m\uparrow}(t')]\rangle &
\langle [c_{m\uparrow}(t), c_{m\downarrow}(t')] \rangle \\
\langle [c^\dagger_{m\downarrow}(t), c^\dagger_{m\uparrow}(t')]\rangle
& \langle [c^\dagger(t)_{m\downarrow}, c_{m\downarrow}(t')] \rangle
\end{matrix}
\right)\nonumber\,.
\end{eqnarray}
The sum over single-particle states $\frac{1}{\nu_S}\sum_{m}$ with
energy $\xi_{m}$ is equivalent to the integration over their energy
$\xi_{\bp}$ in \eref{int_xi}.  Next, we perform the Bogoliubov
transformation from electron operators $c_{m\uparrow}$,
$c_{m\downarrow}$ to quasiparticle operators $\alpha_{m}$,
$\beta_{m}$:
\be
c_{m\uparrow} = u_{m} \alpha_{m} +v_{m} \beta^\dagger_{m} \, ,\quad
c_{m\downarrow} = u_{m} \beta_{m} -v_{m} \alpha^\dagger_{m} \,
\ee
with amplitudes $u_{m}$, $v_{m}$ given by
\be\label{B_am}\begin{split}
&|u_{m}|^2 = 1-|v_{m}|^2=\frac{1}{2}\left(1+\frac{\xi_{m}}{\epsilon_{m}}\right), \\
&u_m v_{m} = -\frac{1}{2}\frac{\Delta}{\epsilon_{m}}
\end{split}\ee
and $\epsilon_{m}=\sqrt{\xi_{m}^2+\Delta^2}$ being the quasiparticle energy.
Introducing the distribution function (we assume equal population of the two spins)
\be\label{fdef}
f(\xi_{m}) = \la \alpha^\dagger_{m} \alpha_{m} \ra = \langle \beta^\dagger_{m} \beta_{m}\rangle \, ,
\ee
using \eref{B_am},
and taking the Fourier transform with respect to $t-t'$ we obtain
\be\label{GKg}\begin{split}
\hG^K_S(\ep) =2\Re \frac{1}{\sqrt{\ep^2-\Delta^2}}
& \Bigl\{ (\ep \tz+i\Delta \ty) [1-2f^E(\eps)] \\ & -
2\sqrt{\ep^2-\Delta^2}f^Q(\eps) \Bigr\}
\end{split}\ee
with the energy and charge modes of the distribution function
$f(\xi_m)$ defined as
\begin{eqnarray}\label{fEQ}
f^E(\eps) = \frac{1}{2}\left[f(\sqrt{\eps^2-\Delta^2}) + f(-\sqrt{\eps^2-\Delta^2})\right],
\nonumber \\
f^Q(\eps) = \frac{1}{2}\left[f(\sqrt{\eps^2-\Delta^2}) - f(-\sqrt{\eps^2-\Delta^2})\right].
\end{eqnarray}
Hence the Keldysh Green's function can be presented in the form
\be
 \hG^K_S = \hG^R_S\,\hat{F} - \hat{F}\,\hG^A_S,
 \label{K-F}
\ee
where
\begin{eqnarray}
\hat F &=& \mathrm{sgn}\,\ep\,[1-2f^E(\eps)]\hat{1}
\nonumber\\
&&-\frac{|\ep|}{\sqrt{\ep^2-\Delta^2}} \left(\tz + i \frac{\Delta}{\ep} \ty\right) 2f^Q(\ep).
\label{Fgen}
\end{eqnarray}
This shows that the trace of $\hat F$ is directly related to the energy mode $f^E$,
while the traceless part possesses an additional density of states factor compared to the
charge mode $f^Q$ of the quasiparticle distribution function.

Comparison of \eref{Fgen} with the solution of the kinetic
equation for $\hat G_S^K$, \eref{K-sol-3},
enables us to find the explicit expressions for the distribution function modes
\begin{eqnarray}
 f^E(\eps) &=& \frac{1}{2}\,
 [1 - \mathrm{sgn}\,\eps\,n^0(\eps)],
 \nonumber\\
 f^Q(\eps) &=& -\frac{1}{2|\eps|}\,
 \sqrt{\eps^2 - \Delta^2}\,
 n^1(\eps)
 \label{fEQto}
\end{eqnarray}
and the voltages $V_L$, $V_R$ are related by \eref{Isum-cond}.
At low temperature and bias $T, eV \ll \Delta$, at leading order in $e^{-\Delta/T}$
we obtain, using \eref{sinh},
\be\label{fE0}
f^{E}(\eps) \approx \Phi(eV/T)\, e^{-\eps/T} \, ,
\ee
with
\be\label{phi-def}\begin{split}
\Phi (eV/T) = \frac{1}{\Gamma_L + \Gamma_R}\sqrt{\Gamma_L^2+\Gamma_R^2 + 2\Gamma_L\Gamma_R\cosh (eV/T)}\, ,
\end{split}\ee
while at this order $f^{Q}$ vanishes due to \eref{cur_cons}.

One may also obtain the distribution function $f_\xi$
by solving the kinetic equation for quasiparticles. The latter method is convenient to assess the effects
of relaxation. In the next section we consider in more detail
the electron-phonon interaction.

\subsection{Kinetic equation for quasiparticles}
\label{sec:ke-qp}

In the kinetic equation approach, the effects of interaction are described by adding
to the rate of change of the occupation numbers due to tunneling an appropriate collision integral $I$.
In the case of electron-phonon interaction, one
can derive a set of coupled kinetic equations for the two modes of the quasiparticle distribution
function, $f^E$ and $f^Q$:
\bea
\frac{d}{dt} f^E & = & \Gamma_L f_{0L}+\Gamma_R f_{0R} - \Gamma f^E  \label{ke-fe}\\  & +&
I^E_r\left\{f^E,f^Q,N\right\} + I^E_s\left\{f^E,f^Q,N\right\} ,\qquad \nonumber \\
\frac{d}{dt} f^Q &
= & -\frac{\sqrt{\eps^2-\Delta^2}}{\eps}
\left[\Gamma_L f_{1L}+\Gamma_R f_{1R}\right]
- \Gamma f^Q
\label{ke-fq}\\ &+& I^Q_r\left\{f^E,f^Q,N\right\} + I^Q_s\left\{f^E,f^Q,N\right\}, \nonumber
\eea
where $\Gamma=\Gamma_L+\Gamma_R$,
%and the terms proportional to $\Gamma$s are due to tunneling
%between superconductor and normal leads.
$f_{0i} = (1 - n^0_i)/2$, and $f_{1i} = n^1_i/2$.
Neglecting the collision integrals, in the steady-state
$df/dt=0$ we recover immediately the result \eref{fEQto}.
The subscripts $r$ and $s$ are used to distinguish, in the electron-phonon
collision integrals, quasiparticle recombination and scattering processes, respectively, and
we take the phonon
distribution function $N$ to be the thermal equilibrium one:
\be
N_\eps = \frac{1}{e^{\eps/T}-1}\, .
\ee

We characterize the electron-phonon collision rate by the scattering
rate, $1/\tau_{ph}$, for normal electrons off phonons at the critical
temperature $T_c$.\cite{foot0} We consider the limit of low
temperature and bias, in which case the electron-phonon interaction
has a small effect on the quasiparticle distribution as long as
tunneling is the main scattering process. Considering relaxation rate
for an arbitrary small perturbation~\cite{scalapino} of the
quasiparticle distribution function, one might conclude that the
corresponding condition for that is $(1/\tau_{ph})(T/\Delta)^{7/2}\ll\Gamma$.
(The small factor $(T/\Delta)^{7/2}$ here suppresses the low
temperature scattering rate for quasiparticles compared to that of
normal electrons at $T_c$.) However, for the specific case of
out-of-equilibrium distribution created by tunneling the condition for
the rate $1/\tau_{ph}$ is even softer, see Eq.~(\ref{condition}) below.

Evaluating the correction to the charge mode $f^Q$ due to the weak
electron-phonon relaxation at low temperatures, we find that such
correction to \eref{fEQto} vanishes at leading order in
$e^{-\Delta/T}$ due to the relation \eref{cur_cons}.

We now consider the energy mode. The collision integrals $I^E_{r,s}$
contain terms quadratic in $f^Q$, which we can neglect since $f^Q$
vanishes at leading order, as discussed above. The non-vanishing terms
are\cite{gal}
\be\label{IEr}\begin{split}
I^E_r = \frac{1}{\tau_{ph}\Delta^3} \int_\Delta^\infty
\!\!d\eps' \,\frac{\eps'}{\sqrt{\eps'^2-\Delta^2}}
\left(\eps+\eps'\right)^2
\left(1+\frac{\Delta^2}{\eps\eps'}\right) \\
\left[
\left(1-f^E_\eps\right)\left(1-f^E_{\eps'}\right)N_{\eps+\eps'} -
f^E_\eps f^E_{\eps'}\left(N_{\eps+\eps'}+1\right)
\right]
\end{split}\ee
for recombination processes and
\be\label{IEs}\begin{split}
& I^E_s = \frac{1}{\tau_{ph}\Delta^3}\bigg\{\int_{\Delta}^{\eps}\!d\eps' \,
\frac{\eps'}{\sqrt{\eps'^2-\Delta^2}} \left(\eps-\eps'\right)^2
\left(1-\frac{\Delta^2}{\eps\eps'}\right) \\ & \left[
\left(1-f^E_\eps\right)f^E_{\eps'}N_{\eps-\eps'} -
f^E_\eps \left(1-f^E_{\eps'}\right)\left(N_{\eps-\eps'}+1\right)
\right] \\
& + \int_{\eps}^\infty \!\!d\eps' \,
\frac{\eps'}{\sqrt{\eps'^2-\Delta^2}} \left(\eps-\eps'\right)^2
\left(1-\frac{\Delta^2}{\eps\eps'}\right) \\ & \left[
\left(1-f^E_\eps\right)f^E_{\eps'}\left(N_{\eps'-\eps}+1\right) -
f^E_\eps \left(1-f^E_{\eps'}\right) N_{\eps'-\eps}
\right]\bigg\}
\end{split}\ee
for scattering processes.

We solve the kinetic equation \rref{ke-fe} by iterations, writing
$f^E$ in the form:
\be\label{fe-it}
f^E \simeq f^{E(0)} + \frac{1}{\tau_{ph}\Gamma} f^{E(1)}
\ee
with $f^{E(0)}$ given in \eref{fE0}.  At leading order in
$e^{-\Delta/T}$, only the scattering collision integral $I^E_s$ is
present. It is satisfied by any Boltzmann distribution function, like
the one in \eref{fE0}.  Indeed, let us consider the terms in square
brackets in the second line of \eref{IEs}; in the low temperature and
bias regime they are approximately
\be
\Phi e^{-\eps'/T}N_{\eps-\eps'} - \Phi e^{-\eps/T}(N_{\eps-\eps'}+1) =
0\,,
\ee
in agreement with the detailed balance in the absence of
recombination.  Similarly, the terms in square brackets in the last
line of \eref{IEs} add up to zero.  Therefore, at order
$e^{-\Delta/T}$ there are no corrections to the distribution function.

To calculate the correction due to recombination, we note that
$\eps +\eps' > 2\Delta$, so that in the last term in square
brackets in \eref{IEr} we can neglect $N_{\eps+\eps'}$ in
comparison to unity. The square brackets then becomes approximately
\be
e^{-(\eps+\eps')/T}\left(1-\Phi^2\right) \, ,
\ee
{\sl i.e.}, they are of order $e^{-2\Delta/T}$. For the correction we find
\be\begin{split}
f^{E(1)} = & \, e^{-\eps/T} \left(1-\Phi^2\right) \frac{1}{\Delta^3} \int_\Delta^\infty\!\!d\eps' \,
\frac{\eps'}{\sqrt{\eps'^2-\Delta^2}}  \\ &\times \left(\eps+\eps'\right)^2
\left(1+\frac{\Delta^2}{\eps\eps'}\right) e^{-\eps'/T}.
\end{split}\ee
After the substitution $\eps' = \Delta + Tx$, the integral can be
evaluated to give, at leading order
\be\label{fe1-f} f^{E(1)} \simeq
e^{-\eps/T} \left(1-\Phi^2\right) e^{-\Delta/T} \sqrt{\frac{\pi
    T}{2\Delta}} \left(\frac{\eps}{\Delta}\right)^2
\left(1+\frac{\Delta}{\eps}\right)^3.
\ee
[Note that at zero bias $\Phi(0) = 1$ and the correction vanishes.] From
this equation we can estimate the ratio between the leading term and
the correction in \eref{fe-it} and find that the iterative solution is
applicable if
\be \label{condition}
\frac{1}{\tau_{ph}\Gamma} \ll
\sqrt{\frac{\Delta}{T}} e^{\Delta/T}\,.
\ee
Under this condition, the electron-phonon relaxation leads only to
a small modification of the distribution function given by
\esref{fe-it} and \rref{fe1-f}.
Albeit the correction Eq.~(\ref{fe1-f}) at low energies
($\eps\approx\Delta$) scales with temperature predominantly as
$e^{-2\Delta/T}$, it may exceed the corresponding term of expansion of
\eref{fEQto} in $e^{-\Delta/T}$. That happens at large values of
$1/\tau_{ph}\Gamma$, which still may satisfy Eq.~(\ref{condition}).

This concludes our analysis of the non-equilibrium steady state of a
superconductor for the specific problem of a finite-bias tunnel
injection of quasiparticles. In the next section we consider the
linear response of an out-of-equilibrium superconductor to an external
AC field.

\section{AC conductivity of N-I-S-I-N structure}
\label{sec:emr}

In this section we study the linear response of the superconductor to an
external AC electric field oscillating at frequency $\omega$. Within linear response, the
Green's function $\cG_S$ is the sum of the zero-approximation isotropic part $\cG_0$
and a small
correction $\cG_1$, linear in ${\bf A}$, which is anisotropic in momentum
space
\be\label{cgstot}
 \cG_S(\bn) = \cG_0 + \cG_1(\bn)\, , \qquad \bn = \bp/|\bp|.
\ee
The correction $\cG_1(\bn)$ determines the current density via
\begin{equation}
 \jw
 =
 -\frac{1}{4}e\nu_S\int\limits_{-\infty}^{\infty}\!\!d\eps\,
 {\rm Tr}\,
 \la {\bf v}\,\tz\,\hG^K_1 \ra,
 \label{j-def}%16
\end{equation}
where ${\bf v}=v_F \bn$, $v_F$ is the Fermi velocity, and angular brackets
denote averaging over the momentum direction.
We emphasize that the calculation of the AC response that we present here is not
restricted to the particular non-equilibrium state considered in the previous section; rather, it is valid
for a generic matrix distribution function $\hat F$, which determines the Keldysh part of the
zero-approximation Green's function $\cG_0$ via \eref{K-F}.

The two terms in the sum \rref{cgstot} obey the orthogonality condition
\be\label{orthcon}
\cG_+ \cG_1(\ep,\omega) + \cG_1(\ep,\omega)\cG_- = 0,
\ee
where
\be\label{cGpm}
\cG_\pm = \cG_0(\ep\pm \omega/2) \, .
\ee

An equation for $\cG_1$ can obtained as before by considering the difference between
\eref{Dyson} and its conjugate and using \esref{int_xi}.
Assuming ${\bf A}(t)=\Aw e^{-i\omega t}$ we find,
at zeroth order in the tunneling rates $\Gamma_i$
\be\begin{split}
 -i\hat H_+\, \cG_1
 +
 i \cG_1\, \hat H_-
 +
 \frac{1}{2\tau}\, \cG_+ \, \cG_1
 -
 \frac{1}{2\tau}\, \cG_1\,  \cG_-
\\ =
 ie{\bf v}\Aw(\tz\,\cG_- - \cG_+\,\tz),
\end{split}\label{r-basic}
\ee
where
\be
\hat H_\pm = \left(\ep\pm\omega/2\right)\tz + i \Delta \ty.
\ee
The zero-order in $\Gamma_i$ approximation is valid when
$\Gamma_L+\Gamma_R \ll \mathrm{min}\{1/\tau,\omega,\Delta\}$.
In particular, the condition $\Gamma_L+\Gamma_R \ll 1/\tau$ implies
that the tunneling has a negligible influence on the quasiparticle states; in fact, the only role
of the leads is to generate a non-equilibrium quasiparticle population, and the calculation
of the AC response does not depend on the specific way in which a non-equilibrium population is
established (as long as it does not substantially alter the quasiparticle states).

Equation \rref{r-basic} can be solved for a generic relation between $1/\tau$  and $\Delta$ --
see Appendix~\ref{sec:app}. Here we consider the dirty limit $\tau\Delta \ll 1$.
To calculate the current, we only need the following expression for the Keldysh component:
\be\begin{split}
& \hG^K_1 =  ie{\bf v}\Aw\tau
\\ & \times \left[
  \hG^R_{+}\,\tz\,(\hG^R_{-} - \hG^A_{-})\,\hat{F}_{-}
  +\hat{F}_{+}\,(\hG^R_{+} - \hG^A_{+})\,\tz\,\hG^A_{-}
 \right]
\end{split}\label{GK-dirty}
\ee
with $\hat{F}_\pm = \hat{F}(\eps\pm\omega/2)$.
The retarded and advanced Green's functions in this equation
are given by \esref{HR-2}-\rref{GR}.
Substituting \eref{GK-dirty} into \eref{j-def},
we arrive at
\be
 \jw  = \sigma(\omega) \mathbf{E}_\omega,
\ee
where $\sigma(\omega)$ is the complex conductivity.
It is convenient to isolate its equilibrium zero-temperature kinetic part
\begin{equation}
  \sigma_0(\omega) = -\frac{\pi \sigma_N \Delta}{i\omega},
  \label{w=0}%26
\end{equation}
which represents the purely inductive response of the
superconducting condensate to the external field in the absence
of quasiparticles and where $\sigma_N =2e^2 \nu_S D$ is the normal-state conductivity of the superconductor, with
$D=v_F^2 \tau/3$ the diffusion constant. Using the relationship \eref{Fgen} between matrix and quasiparticle
distribution functions,
assuming $0<\omega <2\Delta$, and performing simple rearrangements, we find
\be\begin{split}
 & \sigma(\omega)  = \sigma_0(\omega)
 +
 \frac{2\sigma_N}{\omega} \\
& \times \left[\,
  \int\limits_{\Delta}^{\infty} d\ep\,
  \frac
  {\ep\,(\ep +\omega) + \Delta^2}
  {\sqrt{(\ep^2-\Delta^2)\left[(\ep+\omega)^2 - \Delta^2\right]}}\,
  f^E_\ep
 \right.
\\ &
  \quad -
  i\int\limits_{\Delta}^{\Delta+\omega} d\ep\,
  \frac
  {\ep\,(\ep-\omega) + \Delta^2}
  {\sqrt{(\ep^2-\Delta^2)\left[\Delta^2 - (\ep-\omega)^2\right]}}\,
  f^E_\ep
\\ & \ \left.  -
  \int\limits_{\Delta+\omega}^{\infty} d\ep\,
  \frac
  {\ep\,(\ep -\omega) + \Delta^2}
  {\sqrt{(\ep^2-\Delta^2)\left[(\ep-\omega)^2 - \Delta^2\right]}}\,
  f^E_\ep
 \right].
 \label{dj-1}%32
\end{split}\ee
Equation \rref{dj-1} may be viewed as a generalization of the
Mattis-Bardeen formula to an arbitrary quasiparticle distribution
function $f_\xi$.

We now use \eref{dj-1} to find the AC conductivity for the
biased N-I-S-I-N structure considered in Sec.~\ref{sec:steady},
concentrating on the limit of low temperatures and voltages, ${\rm
  max}\{eV_L, eV_R, T\}\ll \Delta$. In this limit, $f^E_\ep$
is exponentially small and approximately given by \eref{fE0}.
If in addition to the assumptions of low temperature and voltages made above, the
frequency is also much smaller than the gap, $\omega \ll \Delta$, we find at
leading order in $\Delta$
\begin{eqnarray}
 \sigma(\omega) =  \frac{\sigma_N\Delta}{i\omega}
 \Bigl\{-\pi + 2 e^{-\Delta/T}\,\Phi(eV/T)
 \nonumber \\
 \times
 \Bigl[
   \pi\exp(-\omega/2T)\,I_0(\omega/2T)
\nonumber\\
  +2i\sinh(\omega/2T)\,K_0(\omega/2T)
 \Bigr]
 \Bigr\},
 \label{j-3}
\end{eqnarray}
where
$I_0$ and $K_0$ denote the modified Bessel functions of zeroth order.

\begin{comment}
Equation \rref{j-3} is our final expression for the conductivity, valid when
$\Gamma_i \ll \omega \ll \Delta \ll 1/\tau$ and $T, eV \ll \Delta$. In the next section we use this
result to study the dependence of the complex conductivity on the density of quasiparticles in the
biased superconductor.
%

\subsection{Quasiparticle density and conductivity}
\end{comment}

The effect of the finite bias on the AC response of the superconductor described by \eref{j-3}
is contained in full in the function $\Phi$, \eref{phi-def}. In fact, as we now show, this function
accounts for the non-equilibrium density of quasiparticles. The quasiparticle density
is given by [cf. \eref{fdef}]
\be
n_{qp}
=2\nu_S\int\!d\xi \, f(\xi)
=
4\nu_S \int_\Delta^\infty\!\!d\eps \,\frac{\eps}{\sqrt{\eps^2-\Delta^2}}
f^E_\eps
\ee
with $f^E$ defined in \eref{fEQ} and the factor 2 accounting for spin.
At low temperature and voltages, we can approximate $f^E$ as
[see \eref{fE0}]
\be
f^E_\ep \simeq \Phi(eV/T) e^{-\ep/T}
\ee
and evaluating the integral at leading order in $\Delta$ we arrive at
\be\label{n-qp}
n_{qp} =4\nu_S \Phi(eV/T) \sqrt{\frac{\pi \Delta T}{2}}e^{-\Delta/T}.
\ee
Using this equation,
\eref{j-3} can be written as
\bea\label{s-nqp}
\sigma(\omega) & = & \sigma'(\omega) + i \sigma''(\omega), \\
\sigma'(\omega) & = & \sigma_N \frac{2\Delta}{\omega}\sinh\frac{\omega}{2T}
K_0\left(\frac{\omega}{2T}\right)\sqrt{\frac{\Delta}{2\pi T}}\frac{n_{qp}}{\nu_S \Delta}, \nonumber \\
\sigma''(\omega) & = & \sigma_N\frac{\pi\Delta}{\omega}\left[
1-
 e^{-\omega/2T}I_0\left(\frac{\omega}{2T}\right)
\sqrt{\frac{\Delta}{2\pi T}}\frac{n_{qp}}{\nu_S \Delta}
\right].\nonumber
\eea
We note that the above result for the AC conductivity in terms of the
quasiparticle density preserves its form even if we include the correction
to the distribution function due to the electron-phonon interaction
[assuming the condition Eq.~(\ref{condition}) is satisfied]. Of course,
the quasiparticle density \eref{n-qp} should be corrected to  account for
\eref{fe1-f}. Equation \rref{s-nqp} agrees with the results obtained in \ocite{Gao} via a phenomenological
generalization of the Mattis-Bardeen formula that includes an effective chemical potential for quasiparticles.

\section{Summary and discussion}

In this work we have derived an extension of the
Mattis-Bardeen formula for the AC conductivity of a dirty
superconductor to include a non-equilibrium occupation of the
quasiparticle states. Equation~\rref{dj-1}
shows that the charge imbalance does not affect
the linear response to external radiation.

As an explicit example, we considered properties of a superconductor
connected via tunnel junctions to two normal leads. Application of a
constant bias to the leads creates a steady non-equlibrium state in
the superconductor, see Sec.~\ref{sec:steady}. We have evaluated
bulk AC conductivity $\sigma(\omega)$ of such N-I-S-I-N setup for low
frequency $\omega\ll \Delta$ of a weak external field and low
temperature ($T\ll\Delta$) while keeping fixed the DC bias $V$.  The
role of the finite bias is to create a non-equilibrium quasiparticle
density $n_{qp}$. Remarkably, it is possible to express the
conductivity $\sigma(\omega)$ as a function of temperature and
$n_{qp}$ only, see \eref{s-nqp}. The dependence of $\sigma(\omega)$ on
bias $V$ enters only through the $V$-dependence of $n_{qp}$. That
simplification works even if electron-phonon relaxation is taken into
account [if the condition \eref{condition} is satisfied]. At zero bias
($V=0$) the distribution $n_{qp}$ becomes the equilibrium distribution
function, and \eref{s-nqp} reduces to the conventional Mattis-Bardeen
formula.

As a further extension one may include the electron-electron
interaction. If sufficiently strong, it would lead to the replacement
of temperature $T$ in the quasiparticle distribution function with an
effective temperature $T_{\mathit{eff}}$.
Interestingly, for a fixed non-equilibrium quasiparticle density, an
increase of temperature (or effective temperature) leads to a lower
dissipation.

\acknowledgments
We are grateful to M.~Devoret, L.~Frunzio, S.M.~Girvin, D.S. Prober,
and R.J.~Schoelkopf for stimulating discussions. KEN thanks Yale
University for the hospitality. His visit was made possible
by the gift of Victor and Marina Vekselberg to Yale University.
This work was supported by IARPA under
ARO Contract No. W911NF-09-1-0369 (LIG) and Yale University (GC).

\appendix
\section{Solution to \eref{r-basic}}
\label{sec:app}

In this Appendix we give some details on how to solve
to \eref{r-basic}. We begin with the equation for $\hG^R_1$
\begin{equation}\begin{split}
 -i\hat H_{+} \hG^R_1
 +
 i\hG^R_1 \hat H_{-}
 +
 \frac{1}{2\tau} \hG^R_{+} \hG^R_1
 -
 \frac{1}{2\tau} \hG^R_1 \hG^R_{-}
 = \\
 -ie{\bf v}\Aw
 ( \hG^R_{+}\tz - \tz\hG^R_{-} ).
 \label{GR1-eq}%2
\end{split}\end{equation}
Expressing $\hat H_{\pm}$ in \eref{GR1-eq} in terms of $\hG^R_{\pm}$ by
means of \eref{GR} and making use of the retarded component
of the orthogonality condition \rref{orthcon}
\be
 \hG^R_{+}\,\hG^R_1 + \hG^R_1\,\hG^R_{-} = 0
\ee
we bring \eref{GR1-eq} to the form
\begin{equation}\begin{split}
 ( -i\xi^R_{+} - i\xi^R_{-} + 1/\tau )\,
 \hG^R_{+}\,\hG^R_1
 = \\
 -ie{\bf v}\Aw\,
 ( \hG^R_{+}\,\tz - \tz\,\hG^R_{-} )
 \label{GR1-eq2}%4
\end{split}\end{equation}
with the obvious notation [cf. \eref{xi-def}]
\be
\xi^R_\pm = [ (\ep\pm\omega/2 + i0^+)^2 - \Delta^2 ]^{1/2}.
\ee
Multiplying both sides of \eref{GR1-eq2} by $\hG^R_{+}$ and making use of
the normalization condition $(G^R_0)^2=1$, we arrive at
\begin{equation}
 \hG^R_1
 =
 \frac{
    e{\bf v}\Aw
    }{
    \xi^R_{+} + \xi^R_{-} + i/\tau
    }\,
    ( \tz - \hG^R_{+}\,\tz\,\hG^R_{-} ).
 \label{GR1-sol}%5
\end{equation}

Consider now the equation for $\hG^K_1$:
\begin{eqnarray}\label{GK1-eq}
 &&-i\hat H_{+} \hG^K_1
 +
 i\hG^K_1 \hat H_{-}
 +
 \frac{1}{2\tau}
 ( \hG^R_{+} \hG^K_1 - \hG^K_1 \hG^A_{-} )
 \\ &&=
  -ie{\bf v}\Aw
 ( \hG^K_{+}\tz - \tz\hG^K_{-} )
 -
 \frac{1}{2\tau}
 ( \hG^K_{+} \hG^A_1 - \hG^R_1 \hG^K_{-} ) \nonumber
\end{eqnarray}
and the corresponding orthogonality condition
\begin{equation}
 \hG^R_{+}\,\hG^K_1 + \hG^K_{+}\,\hG^A_1 + \hG^R_1\,\hG^K_{-} + \hG^K_1\,\hG^A_{-}=0.
 \label{GK1-ortho}%7
\end{equation}
It is convenient to separate $\hG^K_1$ into a ``regular'' part, similar in form to
\eref{K-F}, and an ``anomalous'' one as follows:
\begin{equation}
 \hG^K_1 = \hG^r_1 + \hG^a_1,
 \qquad
 \hG^r_1 = \hG^R_1 \hat{F}_{-} - \hat{F}_{+} \hG^A_1,
 \label{Ga-def}%8
\end{equation}
where $\hat{F}_\pm = \hat{F}(\eps\pm\omega/2)$.
A substitution of this definition into the orthogonality condition
\rref{GK1-ortho} yields the much simpler equation
\begin{equation}
 \hG^R_{+}\,\hG^a_1 + \hG^a_1\,\hG^A_{-} = 0.
 \label{Ga1-ortho}%9
\end{equation}
Substituting \eref{Ga-def} into \eref{GK1-eq} and using
\eref{GR1-eq} we find the equation for $\hG^a_1$
\be\label{Ga1-eq}\begin{split}
&-i\,\hat H_{+}\,\hG^a_1
 +
 i\,\hG^a_1\,\hat H_{-}
 +
 \frac{1}{2\tau}
 ( \hG^R_{+}\,\hG^a_1 - \hG^a_1\,\hG^A_{-} )
 \\
&=
 -ie{\bf v}\Aw
 \left[
  \hG^R_{+}\,(\hat{F}_{+}\,\tz - \tz\,\hat{F}_{-})
  -
  (\hat{F}_{+}\,\tz - \tz\,\hat{F}_{-})\,\hG^A_{-}
 \right].
\end{split}\ee
The solution to this equation can be found by following the procedure similar to that used above to
solve for $\hG_1^R$. First, we
express $\hat H_{+}$ and $\hat H_{-}$ in terms of $\hG^R_{+}$ and
$\hG^A_{-}$, respectively:
\be\label{Ga1-eq1}\begin{split}
& \left(
   -i\xi^R_{+} + \frac{1}{2\tau}
 \right)
 \hG^R_{+}\,\hG^a_1
 -
 \left(
   -i\xi^A_{-} + \frac{1}{2\tau}
 \right)
 \hG^a_1\,\hG^A_{-}
 \\
& =
 -ie{\bf v}\Aw
 \left[
  \hG^R_{+}\,(\hat{F}_{+}\,\tz - \tz\,\hat{F}_{-})
  -
  (\hat{F}_{+}\,\tz - \tz\,\hat{F}_{-})\,\hG^A_{-}
 \right].
\end{split}
\ee
Then, using the orthogonality condition \eref{Ga1-ortho}, we eliminate $\hG^A_-$ in favor of
$\hG^R_+$:
\bea\label{Ga1-eq2}%12
&& ( -i\xi^R_{+} - i\xi^A_{-} + 1/\tau )
 \hG^R_{+}\,\hG^a_1
= -ie{\bf v}\Aw
 \\ && \quad \times\left[
  \hG^R_{+}\,(\hat{F}_{+}\,\tz - \tz\,\hat{F}_{-})
- (\hat{F}_{+}\,\tz - \tz\,\hat{F}_{-})\,\hG^A_{-}
 \right]. \nonumber
\eea
The solution of this equation is obtained by multiplying both sides
by $\hG^R_{+}$ and employing the normalization condition $(\hG^R_0)^2=1$:
\bea \label{Ga1-sol}%13
 \hG^a_1
 &=&
 \frac
 { e{\bf v}\Aw }
 { \xi^R_{+} + \xi^A_{-} + i/\tau }
 \\ &\times& \left[
  \hat{F}_{+}\,\tz - \tz\,\hat{F}_{-}
  -\hG^R_{+}\,(\hat{F}_{+}\,\tz - \tz\,\hat{F}_{-})\,\hG^A_{-}
 \right].
\nonumber
\eea

Equations \rref{GR1-sol} and \rref{Ga1-sol} are valid for $\Gamma_i \ll \Delta, 1/\tau$ and arbitrary
relation between $\Delta$ and $1/\tau$.
In the dirty limit $\Delta \tau \ll 1$, \esref{GR1-sol} and \rref{Ga1-sol} can be
respectively approximated as
\bea
 \hG^R_1 &=& -ie{\bf v}\Aw\tau\,
 ( \tz - \hG^R_{+}\,\tz\,\hG^R_{-} ),
 \label{GR1-dirty}
\\
\label{Ga1-dirty}
 \hG^a_1 &=&  -ie{\bf v}\Aw\tau \\
  &\times & \left[
  \hat{F}_{+}\,\tz - \tz\,\hat{F}_{-}
  -\hG^R_{+}\,(\hat{F}_{+}\,\tz - \tz\,\hat{F}_{-})\,\hG^A_{-}
 \right]. \nonumber
\eea
Substituting these equations into \eref{Ga-def} we find \eref{GK-dirty}.


\begin{thebibliography}{30}
\bibitem{foot0}More precisely, in \esref{IEr}-\rref{IEs} we use for $1/\tau_{ph}$ the definition of \ocite{gal},
which differs from the standard one\cite{scalapino} by the numerical factor $(T_c/\Delta_0)^3$.
Since we are interested in the low-temperature, low-bias regime, we neglect the difference
between the gap $\Delta$ and its zero-temperature, equilibrium value $\Delta_0$.
\end{thebibliography}
\end{document}